\documentclass[]{spie}  %>>> use for US letter paper
\usepackage[T1]{fontenc}
\usepackage{amsmath,amsfonts,amssymb}
\usepackage{upgreek}
\usepackage{array}
\usepackage{graphicx}
\usepackage{aas_macros}
\usepackage[colorlinks=true, allcolors=blue]{hyperref}

\usepackage{xcolor}

\usepackage[inkscapeformat=png]{svg}

\title{Visible Photonic Lantern integration, characterization and on-sky testing on Subaru/SCExAO}

\author[a,b]{S.~Vievard}
\author[a,c]{M.~Lallement}
\author[d]{S.~Leon-Saval}
\author[a,b,e,f]{O.~Guyon}
\author[g]{N.~Jovanovic}
\author[c]{E.~Huby}
\author[c]{S.~Lacour}
\author[a]{J.~Lozi}
\author[a]{V.~Deo}
\author[a]{K.~Ahn}
\author[h]{M.~Lucas}
\author[a,i]{T.~Currie}
\author[j]{S.~Sallum}
\author[k]{M.~P.~Fitzgerald}
\author[d]{C.~Betters}
\author[d,l,m]{B.~Norris}
\author[n]{R.~Amezcua-Correa}
\author[n]{S.~Yerolatsitis}
\author[k]{J. Lin}
\author[k]{Y.~J.~Kim}
\author[g]{P.~Gatkine}
\author[b]{T.~Kotani}
\author[b,o,p]{M.~Tamura}
\author[q]{G. Martin}
\author[c]{H.D.~Kenchington Goldsmith}
\author[c]{G.~Perrin} 

\affil[a]{National Astronomical Observatory of Japan, Subaru Telescope, 650 North Aohoku Place, Hilo, HI 96720, U.S.A. }
\affil[b]{Astrobiology Center, 2-21-1, Osawa, Mitaka, Tokyo, 181-8588, Japan}
\affil[c]{LESIA, Observatoire de Paris, Universit\'e PSL, CNRS, Sorbonne Universit\'e, Universit\'e Paris Cit\'e, 5 place Jules Janssen, 92195 Meudon, France}
\affil[d]{Sydney Astrophotonic Instrumentation Laboratory, School of Physics, The University of Sydney, Sydney, NSW 2006, Australia}
\affil[e]{Steward Observatory, University of Arizona, Tucson, AZ 85721, USA}
\affil[f]{College of Optical Sciences, University of Arizona, Tucson, AZ 85721, U.S.A.} 
\affil[g]{California Institute of Technology, 1200 E California Blvd, Pasadena, CA 91125, U.S.A.}
\affil[h]{Institute for Astronomy, University of Hawai'i, 640 N. Aohoku Pl, Hilo, HI 96720}
\affil[i]{Department of Physics and Astronomy, University of Texas at San Antonio, San Antonio, TX 78006, USA}
\affil[j]{Univ. of California, Irvine, G302 C Student Center, Irvine, CA 92697}
\affil[k]{Univ. of California, Los Angeles, 405 Hilgard Avenue, CA 90095}
\affil[l]{AAO-USyd, School of Physics, University of Sydney 2006}
\affil[m]{Sydney Institute for Astronomy, School of Physics, The University of Sydney, NSW 2006, Australia}
\affil[n]{The College of Optics and Photonics, University of Central Florida, 4304 Scorpius St, Orlando, FL 32816}
\affil[o]{The University of Tokyo, 7-3-1 Hongo, Bunkyo-ku, Tokyo 113-0033, Japan}
\affil[p]{NAOJ, 2-21-1 Osawa, Mitaka, Tokyo 181-8588, Japan}
\affil[l]{Univ. Grenoble Alpes, CNRS, IPAG, 414 Rue de la Piscine, 38400 Saint-Martin-d'Hères, France}

\authorinfo{Further author information: Sebastien Vievard: E-mail: vievard@naoj.org. Based [in part] on data collected at Subaru Telescope, which is operated by the National Astronomical Observatory of Japan.}

\begin{document} 
\maketitle

\begin{abstract}
A Photonic Lantern (PL) is a novel device that efficiently converts a multi-mode fiber into several single-mode fibers. When coupled with an extreme adaptive optics (ExAO) system and a spectrograph, PLs enable high throughput spectroscopy at high angular resolution. The Subaru Coronagraphic Extreme Adaptive Optics (SCExAO) system of the Subaru Telescope recently acquired a PL that converts its multi-mode input into 19 single-mode outputs. The single mode outputs feed a R~4,000 spectrograph optimized for the 600 to 760 nm wavelength range. We present here the integration of the PL on SCExAO, and study the device performance in terms of throughput, field of view, and spectral reconstruction. We also present the first on-sky demonstration of a Visible PL coupled with an ExAO system, showing a significant improvement of x12 in throughput compared to the use of a sole single-mode fiber. This work paves the way towards future high throughput photonics instrumentation at small angular resolution.
\end{abstract}


\section{INTRODUCTION}
\label{sec:intro}  % \label{} allows reference to this section

Astrophotonics is an emerging field that applies photonics technologies to astronomy, including for exoplanet direct imaging and spectroscopy\cite{Currie2023PPVII}. It has enabled high dispersion spectroscopy (HDS) observations of star-planet systems, allowing measurements of orbital velocities, masses and molecular features of giant exoplanets~\cite{snellen10,brogi12}. More broadly, HDS was used to test general relativity via radial velocity measurements of stars in the galactic center~\cite{do19}, characterize stellar populations~\cite{reddy22} or search for intermediate mass black holes~\cite{casares14}.

%Photonics has also been exploited for interferometry, whether for combining the light from several telescopes~\cite{collaboration2017first} or for combining multiple areas of the pupil from a single telescope~\cite{martinod2021scalable,vievard2023singleaperture}. Doing so led to the first direct detection of an exoplanet using interferometry with VLTi/GRAVITY~\cite{lacour2019first}. 

%Single-mode fiber (SMF) fed spectroscopy offers numerous advantages for astronomical observations, compared to multi-mode and/or slit-base spectroscopy. Firstly, it enables achieving higher spectral resolution. Additionally, the compactness and stability of SMF-fed spectrographs enhance spectral quality. However, the main drawback of SMFs lies in the challenge of injecting light into them. This process demands precise alignment of the source to the fiber core and a high Strehl ratio, which can be particularly challenging to achieve, especially in visible light wavelengths.

The advantages of single-mode fiber (SMF) based spectroscopy lie in the compactness and stability of the spectrograph, compared to multi-mode (MM) or slit-based spectroscopy, leading to the development of various SMF-spectrograph instruments. Examples include Subaru/REACH\cite{kotani2020reach}, Keck/KPIC~\cite{delorme2021keck}, and VLT/HiRISE~\cite{vigan2024first}, which aim to inject light into an SMF after correcting for atmospheric turbulence, then disperse the SMF output with high resolution. The "Fizeau" mode of Subaru/FIRST~\cite{vievard2023singleaperture} aims to inject samples of the pupil into SMFs, recombine each sample in free-space, and then disperse the fringes at low or medium resolution. Additionally, Subaru/GLINT~\cite{martinod2021scalable} and the PIC (Photonic Integrated Chip) mode of Subaru/FIRST~\cite{lallement2023photonic} inject the pupil samples into single-mode (SM) waguides engraved into a photonic chip, where the samples are independently recombined and then spectrally dispersed at low or medium spectral resolution.

Extreme Adaptive Optics (ExAO) systems have enabled high injection efficiency in SMFs~\cite{jovanovic2017efficient}, facilitating and encouraging the design of previously mentioned SMF-based spectroscopic instruments. However, while current ExAO systems can achieve up to $95\%$ Strehl in the $H$-band\cite{Currie2023PPVII}, they struggle to reach high Strehl in the visible wavelengths, making it difficult to exploit SMF-based visible spectroscopic instrumentation.

Photonic lanterns (PL)~\cite{leon2010photonic} are waveguide devices that enable high throughput spectroscopy at high angular resolution. A PL efficiently~\cite{birks2015photonic} converts a MM waveguide into several SM waveguides capable of feeding a spectrograph. When coupled with an ExAO system, these devices can significantly enhance on-sky throughput performance compared to SMFs alone - especially in the visible wavelengths where even ExAO systems struggle to reach high Strehl ratios. PLs facilitate the delivery of spectral information at spatial resolutions surpassing those achievable with traditional coronagraphic instruments and are even capable of beating the telescope spatial resolution~\cite{kim2022spectroastrometry}. 

Moreover, PLs are sensitive to both input scene and wavefront. The flux distribution over the single mode waveguides is dependent on the source position/shape~\cite{kim2022spectroastrometry} and on the wavefront input~\cite{norris2020all}. The encoded information delivered by the flux output distribution allows for retrieval of the geometry/position of the observed object, as well as wavefront sensing useful either for instrumental or observable calibration. The latter features are not available when using a sole SMF hence PLs can potentially provide access to a wider range of science applications.

We recently acquired a PL transitioning a MM input into 19~single mode outputs. This PL was installed on the Visible arm of the Subaru Coronagraphic Extreme Adaptive Optics (SCExAO) modular platform at the Subaru Telescope\cite{Jovanovic2015}, as a new injection mode of the FIRST spectro-interferometer. One of the motivations for this upgrade was to explore the PL as a way to increase the overall throughput of the FIRST instrument. The FIRST-Fizeau mode (FIRST-FIZ) recombines 9~sub-apertures, leading to a $24\%$ pupil coverage. Taking into account an average SMF injection efficiency of about $51\%$~\cite{vievard2023singleaperture} without aberrations, the effective light collection from the pupil to the output of the SMFs goes up to $12\%$. FIRST's interest is ultimately to sample the full pupil using micro-lenses to inject into SMFs connected to a photonic chip (FIRST-PIC mode). A full coverage of this mode would lead to an effective light collection of about $37\%$, taking into account pupil coverage ($67\%$), average injection efficiency (goal : $70\%$) and photonic device throughput (goal : $80\%$). The PL would allow to reach $100\%$ of pupil coverage, and should increase the chances of coupling more light into a high throughput photonic device ($> 90\%$~\cite{birks2015photonic}). The PL outputs feed the FIRST medium resolution spectrograph (R~3,000) optimized for wavelengths ranging from about 600~nm to 800~nm. 

In this paper, we first present the integration of the PL on SCExAO (see Section~\ref{sec:integr}). Next, in Section~\ref{sec:charac}, we characterize the device to understand its various working regimes as a function of the changing f-ratio. Finally, Section~\ref{sec:onsky} presents the on-sky testing of the PL, including a comparison with spectroscopy with a sole SMF.

\section{Photonic Lantern spectroscopy on Subaru/SCExAO}
\label{sec:integr}

\subsection{The 19-port photonic lantern}

\begin{figure}[!h]
\centering
	\includegraphics[width=0.9\linewidth]{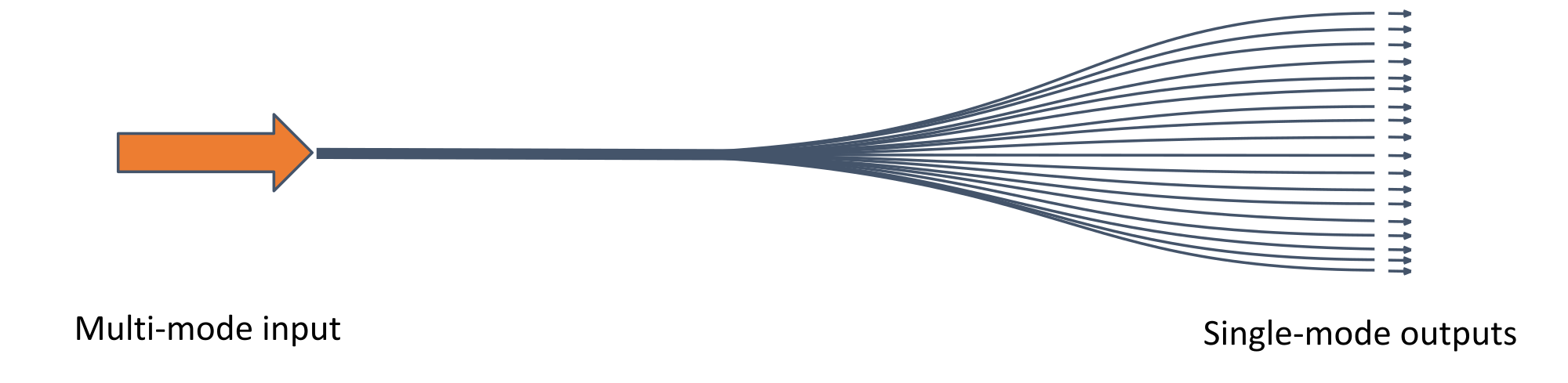}\\
	\caption{Photonic Lantern hardware. The input is a multi-mode fiber, and the outputs are SMFs spliced to a V-groove.}
	\label{fig:pl-input}
\end{figure}

The PL presented in this paper has one MM input and 19~SM outputs (see Figure~\ref{fig:pl-input}). The PL was manufactured using 19 separate SMFs. To create the MM input, one end of the SMFs was inserted into a capillary with a lower refractive index than the SMF fiber cladding. This assembly was then tapered down until the cores of the SMFs disappeared, and their cladding became the new core of the MM region. The 19~pigtailed SMFs are then spliced to a linear V-groove allowing a spectrograph to cross-disperse the light. The transmission of the device was measured by retro-injecting a 780~nm laser and comparing the flux at the SM end and at the MM end. The 19~measures average to $88.9\%$ and span between $83\%$ and $95\%$.

{}

\subsection{Integration on SCExAO}

SCExAO is a versatile platform designed for high-contrast imaging on the Subaru Telescope. It is situated at the Nasmyth Infra-Red focus of the Subaru Telescope, receiving a partially corrected wavefront from the adaptive optics facility AO188~\cite{minowa2010performance}. SCExAO primarily consists of two key components: 1) a pyramid wavefront sensor (PyWFS)~\cite{Lozi_2019} operating in the visible spectrum (approximately $800-950nm$) that delivers wavefront quality exceeding $80\%$ Strehl in the H-band under good conditions, and 2) several coronagraphs that suppress starlight to reveal and study the close circumstellar environment. Over the past decade, SCExAO has grown in complexity and now includes three commissioned science modules~\cite{Currie_2018, vampires, kotani2020reach} as well as several experimental modules, all distributed across two optical benches: an IR bench ($950nm-2.5\mu m$) and a visible bench ($600-950nm$). As shown on Figure~\ref{fig-PL-}, both IR and visible benches share a common path: the light comes either from AO188 or from a calibration source (Super continuum, named SuperK in the following) which can be inserted. The beam is then collimated by an off-axis parabola (OAP) toward a Boston Micromachine 2,000 actuator mirror driven by the PyWFS for wavefront correction. Before being split by a dichroic to separate visible and IR wavelengths, the beam passes through a pupil mask mimicking the Subaru Telescope pupil for calibration purposes. 

\subsubsection{A new mode to the FIRST injection module}
\label{sec:inj_module}

%The FIRST instrument at Subaru~\cite{vievard2023singleaperture} injects stellar light at visible wavelengths in the pupil plane before interferometric recombination. The motivations for using a visible photonic lantern are quite similar to the ones of FIRST. The difference is that for a PL the interferometric recombination happens in a multimode cavity (the multimode fiber area), and that the stellar light injection happens inside the focal plane. It was thus logical to upgrade the FIRST instrument with photonic lantern capabilities by adding the FIRST injection bench with a new injection module in the focal plane. 

\begin{figure}[!h]
\centering
	\includegraphics[width=0.9\linewidth]{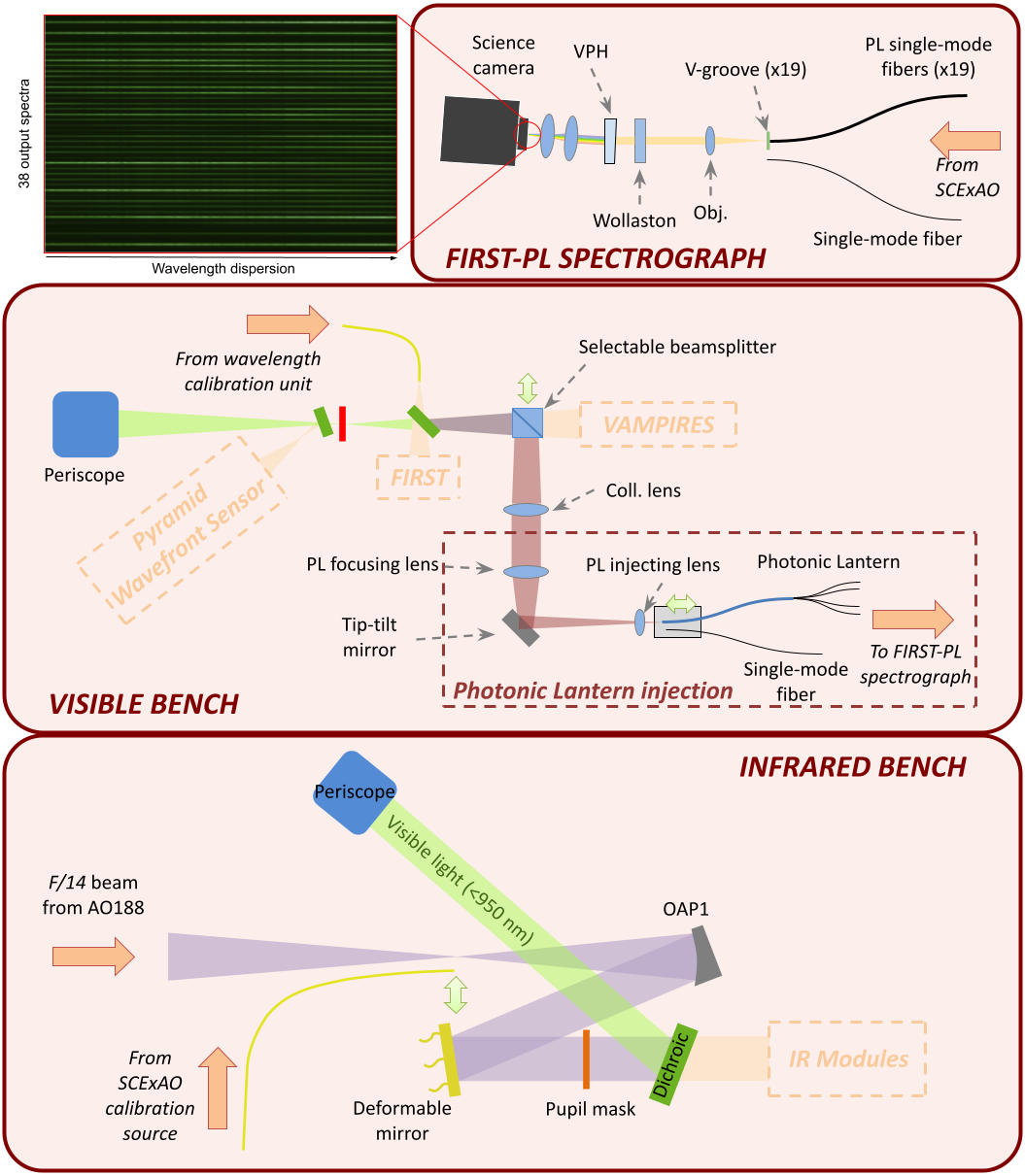}\\
	\caption{Photonic Lantern integration on the SCExAO Visible bench.}
	\label{fig-PL-}
\end{figure}

\begin{figure}[!h]
	\centering
	\includegraphics[width=\linewidth]{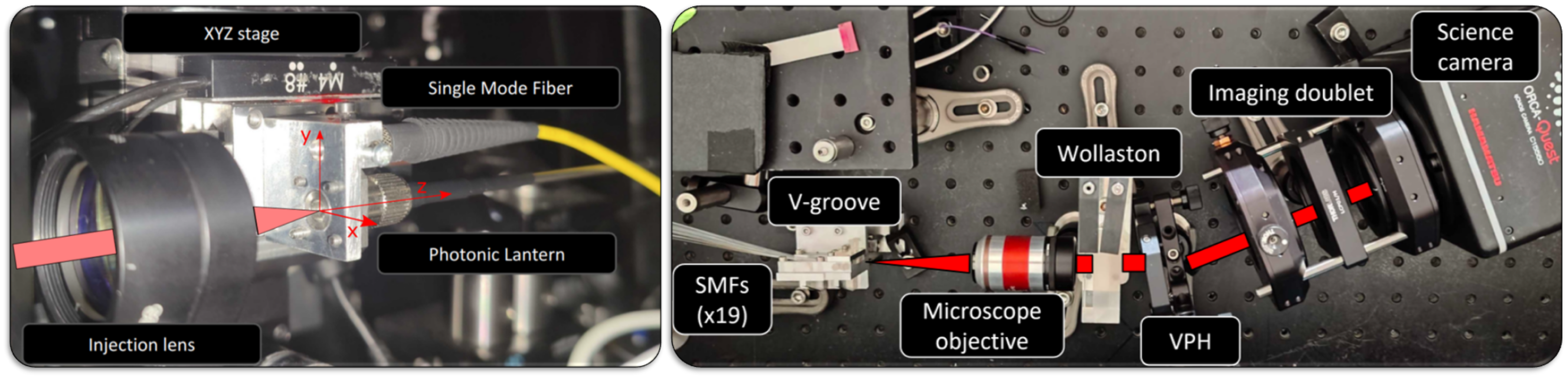}\\
	\caption{Left: Image of the injection module, which includes the injection lens, the PL, and the SMF used for calibration. Right: Image of the R 3000 spectrograph. This spectrograph, fed by the SMF, disperses light using a volume phase holographic grating and enables polarization splitting of the signal with the help of a Wollaston prism.}
	\label{fig-PL-inj}
\end{figure}

The purpose of this new injection module is to efficiently inject light into the PL. The following experimental sections will show that there are several ways to inject light into the PL. The main parameter to be tuned is the input focal ratio. Figure~\ref{fig-PL-} shows the optical layout of the PL injection. The incoming beam from SCExAO has a focal ratio of f/28.4, and is picked up by a 50/50 beamsplitter cube - allowing simultaneous imaging using the VAMPIRES instrument. A collection of 2 to 3 lenses are used to adapt the f-ratio and inject light into the PL. The configuration showed on Figure~\ref{fig-PL-} corresponds to a focal ratio of f/4. The PL is mounted on a bracket which can be remotely moved on 3~axis as shown on Fig.~\ref{fig-PL-inj}. For calibration purposes, an SMF is also mounted on the bracket. 

\subsubsection{The mid-resolution spectrograph}
\label{sec:inj_module}

The spectrograph is designed to be fed by either SMFs positioned in a V-groove with a pitch of 127~$\mu m$, or by the sole SMF. A 2x apochromatic microscope objective collimates the fiber output beams before they are dispersed by a volume phase holographic (VPH) grating from Wasatch Photonics. Two achromatic doublets then image the fiber outputs onto a Hamamatsu ORCA-Quest camera, achieving a resolution of about 3000. In addition, and prior to the VPH, we added a wollaston in order to split the 2~polarizations.

{}

{}

\section{Photonic Lantern lab characterization}
\label{sec:charac}
This section focuses on analyzing the coupling of light into the photonic lantern at various f-ratios. 

\subsection{Coupling light into the photonic lantern}
\begin{figure}[!h]
\centering
	\includegraphics[width=\linewidth]{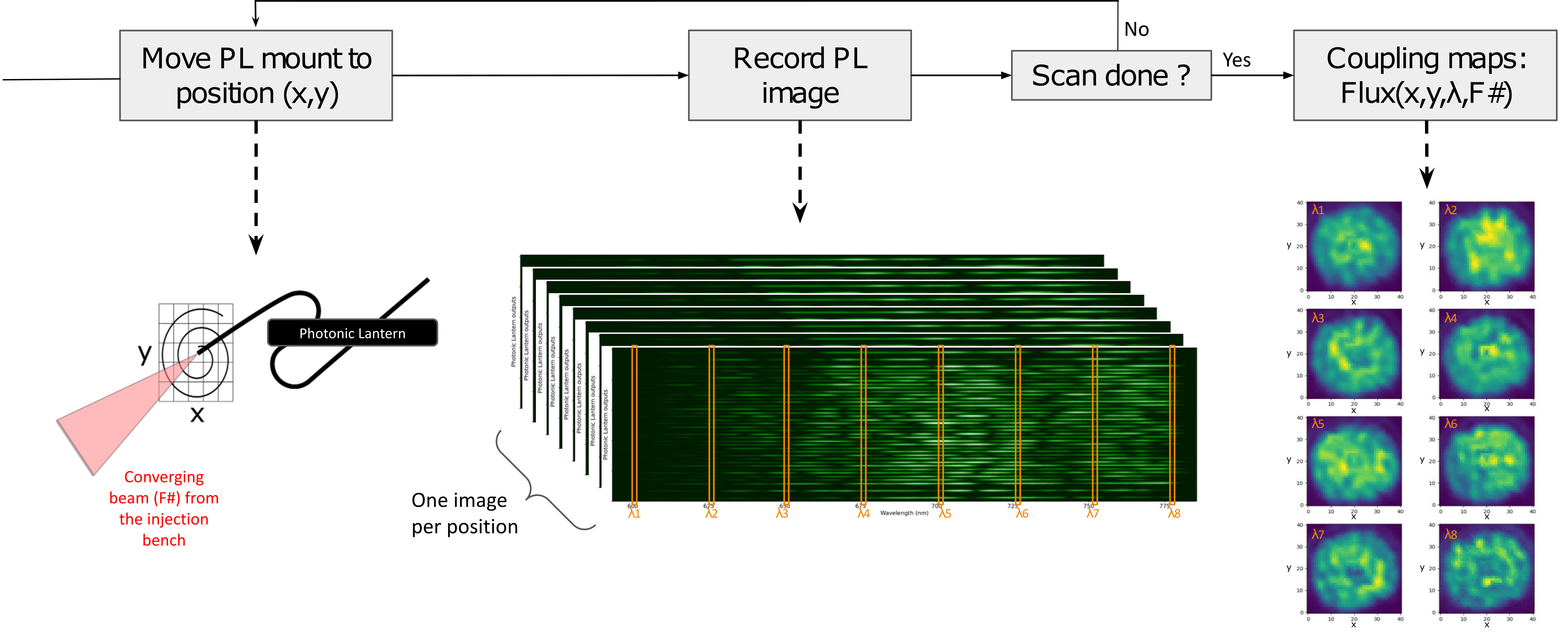}\\
	\caption{Injection optimization procedure. The PL is scanned in the focal plane. For each scanned position, an image is acquired and the total flux per spectral channel is extracted to build coupling maps. The coupling maps are use to position the PL on the optical axis.}
	\label{fig-PL-inj-proc}
\end{figure}

The initial step is to properly inject light into the PL. Figure~\ref{fig-PL-inj-proc} illustrates our procedure. The PL stage is translated in the (X,Y) plane to scan the focal plane. At each scanned position, an image is captured, and the total flux per spectral channel is recorded. The source used for this test is the SCExAO SuperK. After completing the scan, we reconstruct coupling maps, which display the total flux recorded at the output of the PL for each (X,Y) position for each wavelength. We then perform a 2D gaussian fit in order to position the PL at the (X,Y) position corresponding to the center of the coupling map, corresponding to the optical axis. 

{}

\subsection{Coupling maps vs focal ratio}
\begin{figure}[!h]
	\centering
	\includegraphics[width=0.75\linewidth]{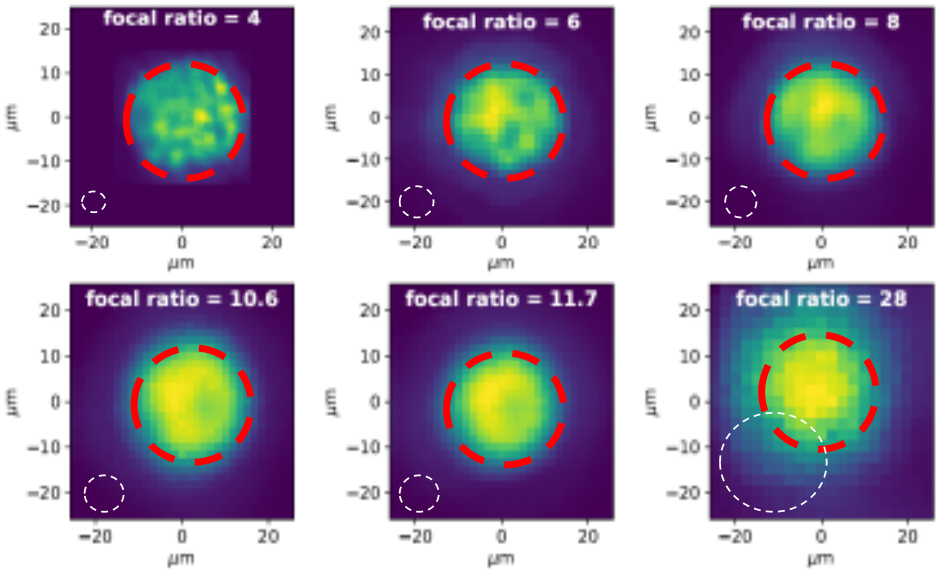}\\
	\caption{Coupling maps acquired for various focal ratios at a fixed wavelength of 765.5 nm.}
	\label{fig-couplmaps}
\end{figure}

We obtained several coupling maps at various focal ratios, as shown in Fig.~\ref{fig-couplmaps}, using the SCExAO SuperK. Each coupling map includes the mode field diameter size (approximately 25~$\mu m$) marked in red and the PSF size in white. It is evident that the coupling maps become more granular as the f-ratio decreases. This granularity occurs because the PSF becomes smaller relative to the MFD as the f-ratio decreases.

{}

\subsection{Spectrum reconstruction vs focal ratio}
\begin{figure}[!h]
	\centering
	\includegraphics[width=0.65\linewidth]{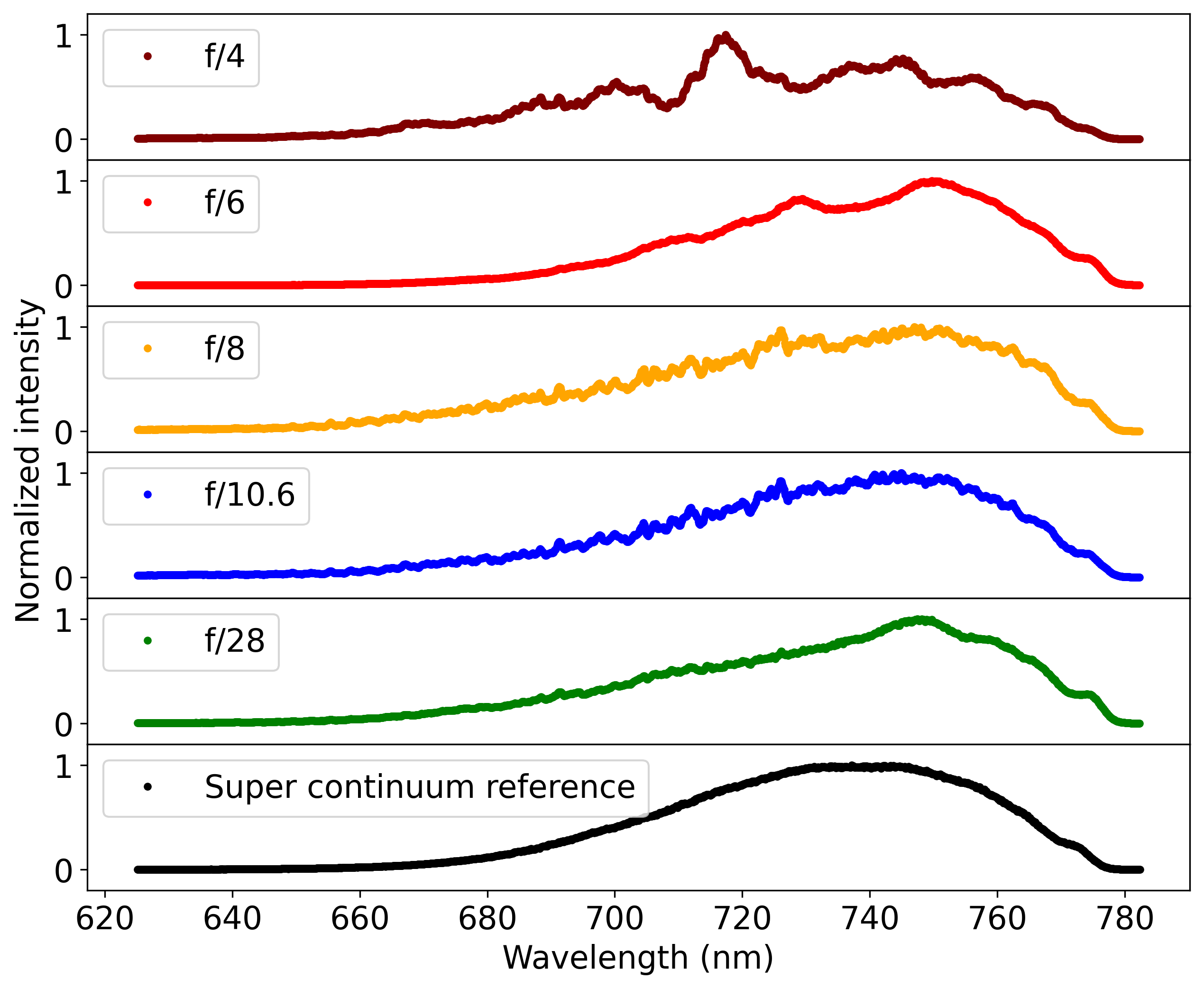}\\
	\caption{Reconstructed supercontinuum spectra from PL data at various focal ratios. Each spectrum is normalized to its maximum value. These spectra are compared to the reference spectrum at the bottom, obtained with an SMF.}
	\label{fig-specrecons}
\end{figure}

We investigated spectrum reconstruction for various f-ratios. The test was realized using the SCExAO SuperK. The reconstruction procedure involves extracting the 38 traces from the image and then co-adding them. The results are shown in Figure~\ref{fig-specrecons}. A reference spectrum was obtained using the SMF. When examining the reconstructed spectra for different f-ratios, spectral drops or low-frequency patterns are noticeable. These spectral drops occur due to modal overfilling of the PL when the f-ratio is too small.

{}

\subsection{Injection efficiency vs focal ratio}
\label{sec-pleff}
For each tested focal ratio, we optimized the injection by creating a coupling map with the SCExAO SuperK and positioning the PL at the map's estimated center. After optimization, we switched the source to a 642 nm laser and recorded the flux intensity before ($I_b$) and after ($I_a$) the PL using a power meter. We normalized out the Fresnel reflections ($96\%$ per surface) and the average PL throughput ($88\%$) to compute the coupling injection efficiency ($c_\text{eff}$) using the formula: $c_\text{eff} = I_a / (I_b \times f_{r}^2 \times t_p)$. Due to granulation in the coupling maps, efficiency varies across the FoV. We converted the on-axis measurement at 642 nm into a coupling efficiency map and computed the maximum and averaged injection efficiencies over the FoV.

\begin{figure}[!h]
	\centering
	\includegraphics[width=\linewidth]{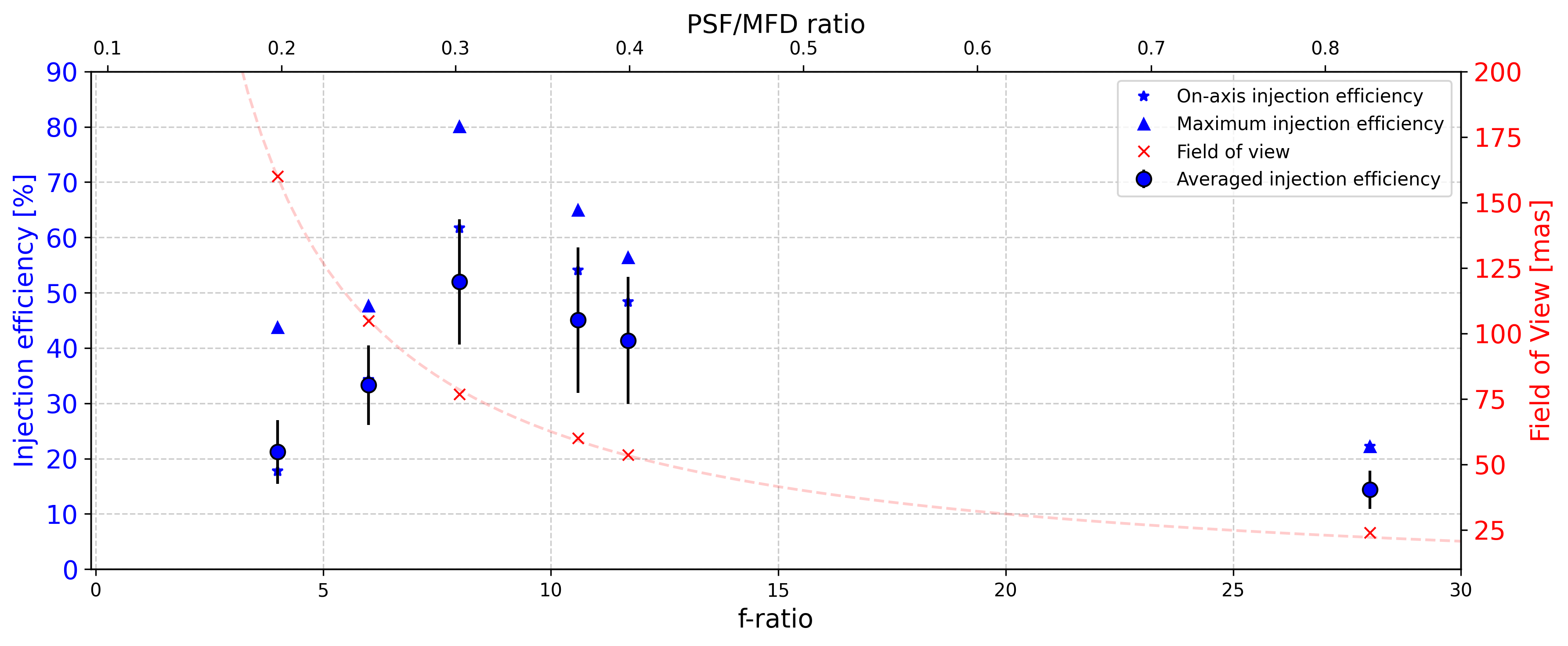}\\
	\caption{Variation of the injection efficiency and the field of view as a function of the focal ratio. The measurements were made using a 642 nm laser.}
	\label{fig-injeff}
\end{figure}

Figure~\ref{fig-injeff} summarizes our tests, showing on-axis, maximum, and average injection efficiencies for various focal ratios. The error bars represent the standard deviation of efficiency across the FoV. Injection efficiency increases rapidly from focal ratios 4 to 8, peaking at 8 with an average efficiency of $51\% \pm 10\%$ over a 80 mas FoV, an on-axis efficiency of $61.8\%$, and a maximum efficiency of $80\%$. Efficiency decreases for focal ratios from 8 to 28, a trend observed by in similar studies\cite{lin2022experimental}.

The graph also shows the FoV evolution as a function of focal ratio, following a hyperbolic curve, as the fiber's acceptance angle is inversely proportional to the focal ratio. This information helps determine the optimal PL injection focal ratio based on the scientific application. Lower focal ratios provide a larger FoV but lower efficiency, with potential spectral information loss. Higher focal ratios offer better spectral reconstruction but a smaller FoV.

{}

\section{Photonic Lantern on-sky testing}
\label{sec:onsky}

We present in this section two on-sky tests. We first present the observation of a point source, Ikiiki ($\alpha$ Leo), using a f-ratio providing optimal injection efficiency into the PL. We then present the observation of a triple system, Po'a (Algol) both with the PL and the SMF mounted on the injection setup. The triple system presents an unresolved binary and a component outside of the PL FoV. The selected f-ratio for this second test was 4, as it is optimal for injection into the SMF, according to the manufacturer specifications. We only present here preliminary results on this target, by comparing the acquisition with the PL and the SMF.

\subsection{Observation of Ikiiki}
We observed Ikiiki ($\alpha$ Leo, $R_{mag}$ = 1.37) on February 18th, 2024 UTC during a telescope Engineering night. The f-ratio selected for this observation was 8, as it showed to provide optimal injection efficiency into the PL, from the tests presented in Section~\ref{sec-pleff}.

\begin{figure}[!h]
	\centering
	\includegraphics[width=\linewidth]{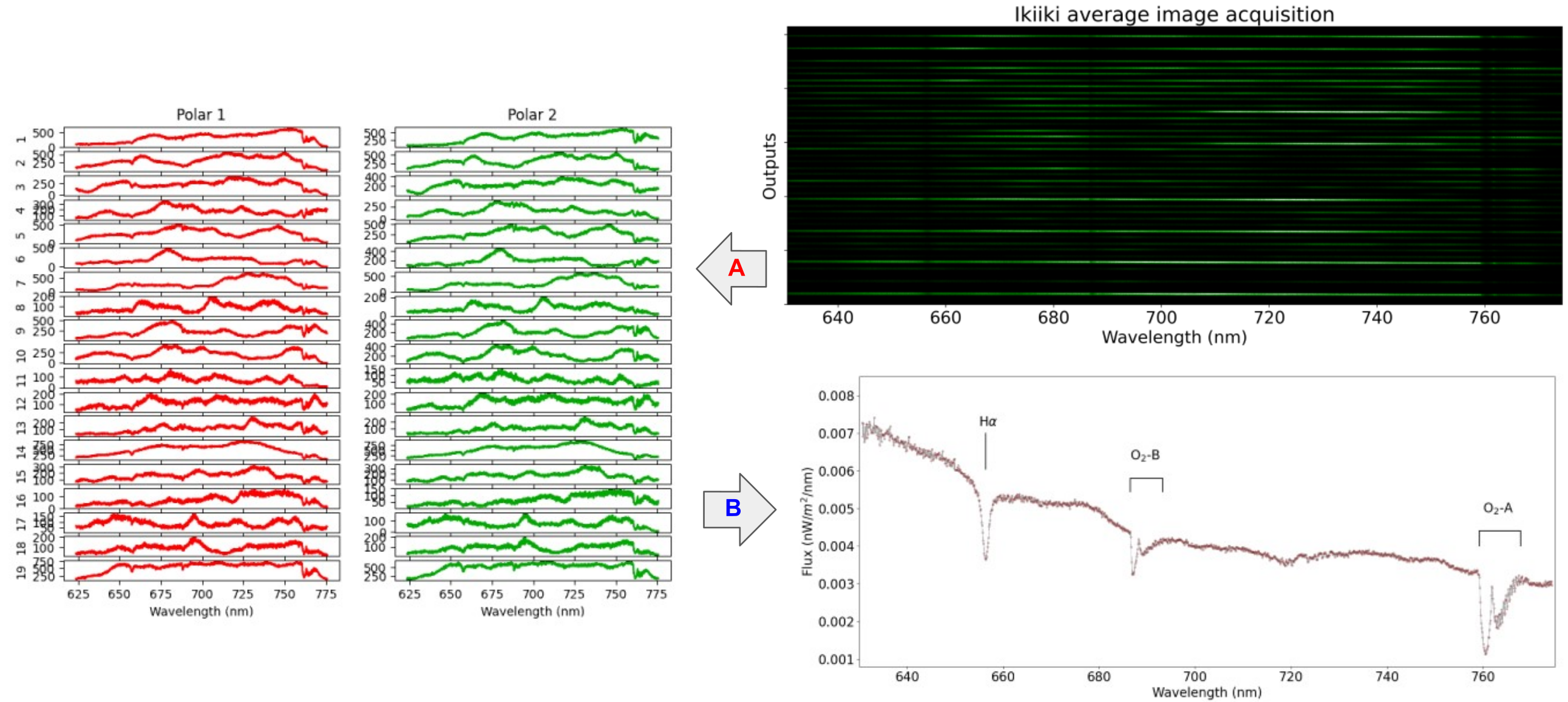}\\
	\caption{Observation of Ikiiki. The spectra of each polarization from each output are extracted (step A), then summed and calibrated using a flat field halogen lamp (step B).}
	\label{fig-ikiiki}
\end{figure}

We observed Ikiiki for 20 minutes at a frame rate of 200 Hz, resulting in a total of 240,000 frames. Due to suboptimal observing conditions, not all the frames could be utilized. We selected the best $90\%$ of the frames based on the flux received by the PL science camera. The top right image on Figure~\ref{fig-ikiiki} shows a dark-subtracted averaged image after this frame selection process. In this data, we can already identify three prominent absorption lines. 

For each selected frame, we subtracted the dark and bias, then extracted (step A on Figure~\ref{fig-ikiiki}) each PL output trace. These traces were averaged over the entire observation sequence. To calibrate our instrument's response, we used a flat field halogen lamp (approximate black body spectrum) at the entrance of AO188 to uniformly illuminate the PL input FoV. By following the same (A) computation method, we derived a flat field spectrum per output, which we calibrated using the theoretical black body slope for the halogen lamp temperature (3020.3K). This calibrated flat data was then used to calibrate the Ikiiki extracted spectra obtained with the PL.

Finally, we co-added all the output spectra and refined the resulting spectrum by normalizing it to match a reference spectrum~\cite{1996BaltA...5..603A} (step B on Figure~\ref{fig-ikiiki}). The spectrum of Ikiiki shows three distinct absorption lines: Atmospheric oxygen A and B bands (around 761 nm and 687 nm, respectively) and H$\alpha$ (approximately 656 nm). The oxygen bands result from oxygen molecules in Earth's atmosphere and are absent in the reference spectrum due to its terrestrial origin. At a spectral resolution of R~3000, the $O_2$ absorption bands appear as a deep, broad line with a forest of narrower lines at longer wavelengths. The H$\alpha$ absorption line originates from the cooler outer layers of the Ikiiki atmosphere, where neutral hydrogen atoms absorb photons from the deeper, hotter layers at the energy level corresponding to the H$\alpha$ transition.

{}

%\begin{figure}
%    \centering
%    \begin{tabular}{cc}
%     \includegraphics[width=0.45\linewidth]{20240218Regulus_flux_tot.png}  &
%     \includegraphics[width=0.45\linewidth]{Regulus_injection_eff.png}
%    \end{tabular}
%    \caption{Caption}
%    \label{fig:enter-label}
%\end{figure}

\subsection{Photonic Lantern versus single-mode fiber spectroscopy on Po'a}

We observed Po'a (Algol, $V_{mag}$ = 2.12) on August 30th, 2023 UTC during a telescope engineering night. The goal of this observation was to compare the performance of spectroscopy using the PL and the SMF. The f-ratio was set to 4 to match the numerical aperture of the SMF (provided by Thorlabs, about 0.13). We acquired 3000 frames, each with a 20 ms exposure, for both PL and SMF cases.

\begin{figure}
    \centering
    \includegraphics[width=\linewidth]{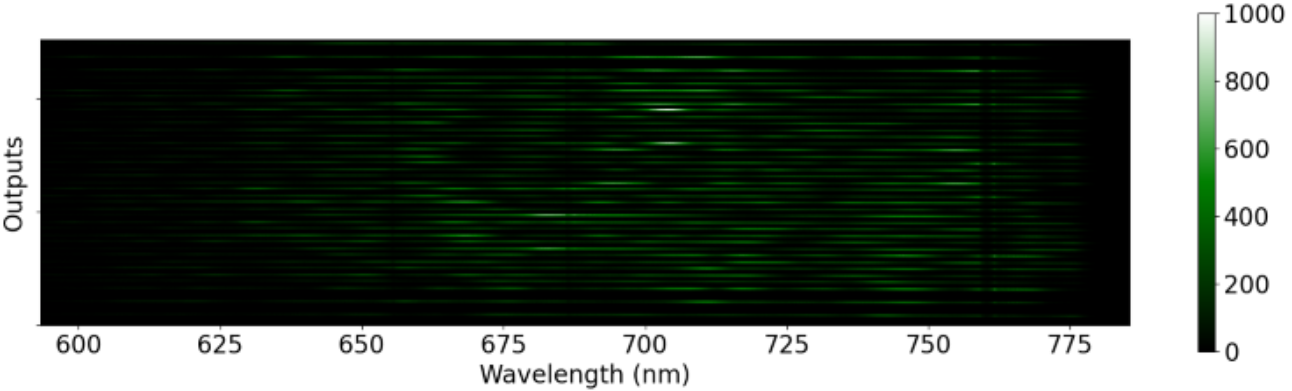} \\
    \includegraphics[width=\linewidth]{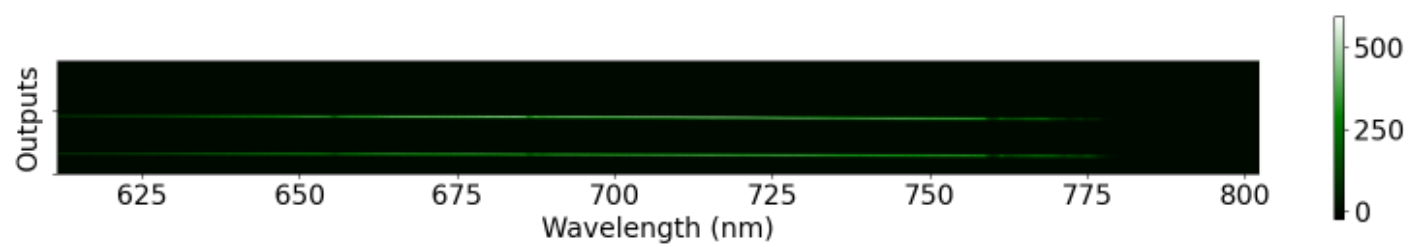} \\
    \includegraphics[width=\linewidth]{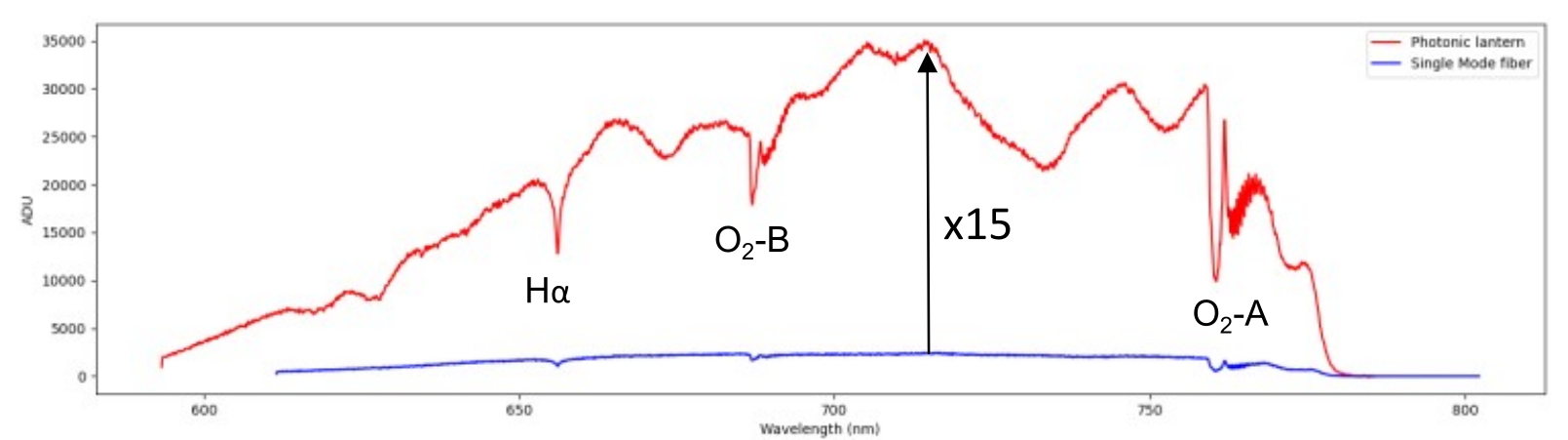} \\
    \caption{Top: PL acquisition of Po'a averaged from 3000 frames of 20 ms each. Middle: SMF acquisition of Po'a averaged from 3000 frames of 20 ms each. Bottom: Reconstruction of Po'a spectrum from the PL (red) and the SMF (blue) data.}
    \label{fig:enter-label}
\end{figure}

Figure~\ref{fig:enter-label} shows the dark-subtracted, averaged images (each averaged over 3000 frames) for the PL (top) and the SMF (middle), using the same color scale for both. The PL image contains the 38 traces resulting from both polarizations of each PL output while the SMF image contains both polarizations of the SMF output. The image scales show higher counts on the PL image, intuiting a better light collection than with the SMF. 

In both cases, we extracted the traces and summed them to obtain a reconstructed spectrum of Po'a. The graph on the bottom of Figure~\ref{fig:enter-label} shows the counts in ADUs of the reconstructed spectra as a function of wavelength. The red spectrum, reconstructed from the PL data, shows spectral drops caused by the low f-ratio, but the H$\alpha$, $O_2$-A, and $O_2$-B absorption lines are still distinguishable. The blue spectrum is the reconstructed spectrum using the SMF data, and shows the same three absorption lines. A comparison of the flux recorded in each spectral channel reveals that the PL data contains, on average, 12.5 times more flux than the SMF data, with a peak gain of 15 times around 720~nm.

We can illustrate this difference in light collection using the VAMPIRES data. VAMPIRES is an instrument located on the SCExAO visible bench (see Figure~\ref{fig-PL-}), operating in transmission after the PL injection beamsplitter cube. We used a 50/50 beamsplitter cube to record images using the focal plane camera of the VAMPIRES instrument. Images were recorded simultaneously during the PL and SMF tests, with a bandwidth spanning from 693 to 744 nm, and a central wavelength at 719 nm. Details about the new VAMPIRES setup will be available soon in a paper by M. Lucas~\cite{miles_vamps}.

\begin{figure}
    \centering
    \includegraphics[width=\linewidth]{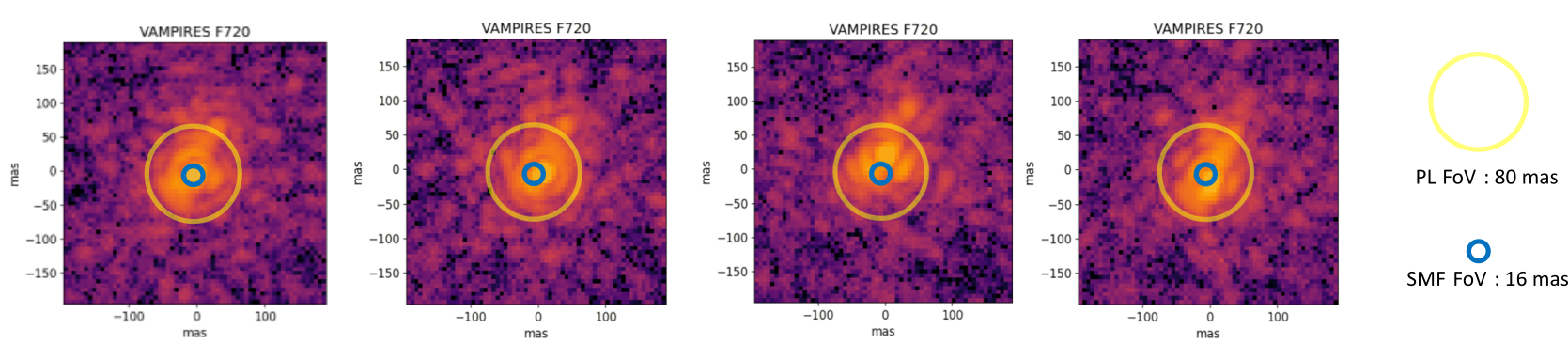}
    \caption{time series of Po'a observation acquired simultaneously with the PL and SMF tests. We overlaid the FoV of the PL (yellow) and the SMF (blue).}
    \label{fig:vampires}
\end{figure}

Figure~\ref{fig:vampires} shows a time series of the VAMPIRES recorded images. Overlaid on the focal plane images are the fields of view of both the PL (in yellow, 80 mas FoV) and the SMF (in blue, 16 mas FoV). From one image to the next, we can see that residual tip-tilt prevents the core part of the PSF from being consistently centered on the SMF FoV, while it remains within the FoV of the PL. Additionally, low-order aberrations distort the PSF, preventing optimal injection into the SMF while the general shape of the PSF stays within the PL FoV. Finally, high-order aberrations send light outside both the PL and SMF FoVs.

This highlights the advantage of using a PL for SMF-fed spectroscopy compared to using an SMF alone, especially under suboptimal atmospheric conditions and/or ExAO correction. The multi-mode input of the PL facilitates easier light injection in the presence of residual tip/tilt and low-order aberrations. A future paper will provide a quantified comparison between PL and SMF spectroscopy.

\section{Conclusion}

We have presented the integration, characterization, and on-sky performance of a photonic lantern at the 8.2-meter Subaru telescope. Photonic lanterns enable highly efficient SMF-based spectroscopy due to their particular design. Their multi-mode input ensures superior coupling efficiency compared to a single SMF, particularly at visible wavelengths where even extreme adaptive optics (ExAO) systems face challenges in achieving high Strehl ratios. The SM outputs of photonic lanterns are ideally suited for use with compact and stable spectroscopic instruments.

We have characterize our visible 19~modes PL, installed on the SCExAO platform at the Subaru Telescope. We confronted the focal ratio of the injection setup with the injection efficiency and the spectrum reconstruction. Our setup achieves optimal injection efficiency at a focal ratio of approximately 8, with an average efficiency of around $51\% \pm 10\%$ over the PL FoV, and a peak efficiency of $80\%$ on-axis. The reconstruction of the SCExAO internal SuperK source presented spectral drops for decreasing focal ratios. These spectral drops were attributed to a modal overfilling of the PL, where some modes at various wavelengths are not transmitted by the PL. 

Finally, we tested the PL on-sky by observing two targets: Ikiiki ($\alpha$ Leo) and Po'a (Algol). The observation of Ikiiki confirmed our ability to obtain, on-sky, PL data and reconstruct the spectrum, retrieving various expected spectral absorption lines. Further studies on this data set will be published in an upcoming paper. The observation of Po'a allowed us to compare the efficiency of the PL to that of an SMF. Under median seeing conditions, and with a focal ratio optimized for the SMF but not for the PL, we found that the PL data contained 12 times more flux than the SMF data. Due to its multi-mode input, the PL has a larger field of view compared to the SMF, allowing it to capture more light in the presence of low-order aberrations and/or residual tip/tilt. A more quantitative study of this comparison will be published in an upcoming paper.

This work paves the way towards future high throughput photonics instrumentation at small angular resolution. Benefiting from the large collecting area and high resolution power offered by extremely large telescopes would bring these instruments closer to potentially discovering and characterizing Earth-like exoplanets.

% Notes : Justify using FIRST-PL  : increase throughput. FIRST-FIZ with 9 sub-aps : $24\%$ pupil coverage, leading to $12\%$ when taking into account the average injection efficiency in the SMFs~\cite{vievard2023singleaperture}. The interest of FIRST is ultimately to exploit the full pupil. Extrapolating FIRST-FIZ to the full pupil would lead to the sampling of 30~sub-aps, which would correspond to a $67\%$ pupil coverage ratio. If we keep the same average injection efficiency, it leads to a $34\%$ effective light collection. Note that developments are in progress in order to increase the injection efficiency into the SMFs (try to find citations) : 3D-printed microlenses on the SMF input. The PL gives the advantage of collecting $100\%$ of the pupil. considering the $60\%$ on-axis injection efficiency obtained during previous tests (cite) leads to a $53\%$ effective pupil light collection.
%\newpage

%\begin{table}[!h]
%\centering
%\begin{tabular}{|c|c|c|c|}
%\hline
%    FIRST Mode &  FIZEAU-9 & PIC-30 & PHOTONIC LANTERN \\
%\hline
%    Pupil coverage & \includegraphics[width=0.2\linewidth]{FIRST-FIZ_pup.png} & \includegraphics[width=0.2\linewidth]{FIRST-FIZ-ideal_pup.png} &  \includegraphics[width=0.2\linewidth]{FIRST-PL_pup.png}\\
%\hline
%    Pupil coverage ratio & 20\% & 67\% & 100\% \\
%\hline
%    Average injection efficiency &  51\% & 70\% & 60\% \\
%\hline
%    Photonics throughtput & NA & 80\% & 89\% \\
%\hline
%    Effective light collection & $10\%$ & 37\%& 53\% \\
%\hline
%\end{tabular}
%\end{table}

\acknowledgments % equivalent to \section*{ACKNOWLEDGMENTS} 
The development of FIRST was supported by Centre National de la Recherche Scientifique CNRS (Grant ERC LITHIUM - STG - 639248), by the French National Research Agency (ANR-21-CE31-0005) and the Action Spécifique Haute Résolution Angulaire (ASHRA) of CNRS/INSU co-funded by CNES. M.L acknowledges support from the doctoral school Astronomy and Astrophysics of Ile de France (ED 127) and from the ANR-21-CE31-0005. S.L. also acknowledges the support of the ANR, under grants ANR-21-CE31-0017 (project ExoVLTI) and ANR-22-EXOR-0005 (PEPR Origins).
The development of SCExAO was supported by the Japan Society for the Promotion of Science (Grant-in-Aid for Research \#23340051, \#26220704, \#23103002, \#19H00703 \& \#19H00695), the Astrobiology Center of the National Institutes of Natural Sciences, Japan, the Mt Cuba Foundation and the director’s contingency fund at Subaru Telescope. This work was supported by the National Science Foundation under Grant No.s 2109231 \& 2308360.
K.A. acknowledges funding from the Heising-Simons foundation.
V.D. acknowledges support from NASA funding (Grant \#80NSSC19K0336).
M.T. is supported by JSPS KAKENHI grant Nos.18H05442, 15H02063, and 22000005.

The authors wish to recognize and acknowledge the very significant cultural role and reverence that the summit of Mauna Kea has always had within indigenous Hawaiian communities, and are most fortunate to have the opportunity to conduct observations from this mountain.

% References
\bibliography{main} % bibliography data in report.bib
\bibliographystyle{spiebib} % makes bibtex use spiebib.bst

\end{document}